\documentclass[preprint,12pt]{aastex}
\usepackage{epsfig}
\usepackage{emulateapj5}
\usepackage{apjfonts}

\newcommand{\chandra}{{\it Chandra}}

\newcommand{\rosat}{{\it ROSAT}}

\newcommand{\einstein}{{\it Einstein}}
\newcommand{\lum}{\thinspace\hbox{$\hbox{ergs}\thinspace\hbox{s}^{-1}$}}
\newcommand{\flux}{\thinspace\hbox{$\hbox{ergs}\thinspace\hbox{cm}^{-2}\thinspace\hbox{s}^{-1}$}}

\begin{document}

\def\spose#1{\hbox to 0pt{#1\hss}}
\def\laeq{\mathrel{\spose{\lower 3pt\hbox{$\mathchar"218$}}
     \raise 2.0pt\hbox{$\mathchar"13C$}}}
\def\gaeq{\mathrel{\spose{\lower 3pt\hbox{$\mathchar"218$}}
     \raise 2.0pt\hbox{$\mathchar"13E$}}}

\makeatletter
\newenvironment{inlinetable}{%
\def\@captype{table}%
\noindent\begin{minipage}{0.999\linewidth}\begin{center}\footnotesize}
{\end{center}\end{minipage}\smallskip}

\newenvironment{inlinefigure}{%
\def\@captype{figure}%
\noindent\begin{minipage}{0.999\linewidth}\begin{center}}
{\end{center}\end{minipage}\smallskip}
\makeatother

\slugcomment{Accepted for publication in the Astronomical Journal}

\title{X-ray/Optical/Radio Observations of a Resolved Supernova Remnant in
NGC~6822} 

\author{Albert~K.~H.~Kong}

\affil{Harvard-Smithsonian Center for Astrophysics, 60 Garden Street,
Cambridge, MA 02138; akong@cfa.harvard.edu}

\author{Lor\'ant~O.~Sjouwerman}
\affil{National Radio Astronomy Observatory, Socorro, NM 87801}

\author{and}

\author{Benjamin~F.~Williams}
\affil{Harvard-Smithsonian Center for Astrophysics, 60 Garden Street,
Cambridge, MA 02138}

\label{firstpage}

\begin{abstract}
The supernova remnant (SNR), Ho~12, in the center of the dwarf
irregular galaxy NGC~6822 was previously observed at X-ray, optical,
and radio wavelengths. By using archival \chandra\ and ground-based
optical data, 
we found that the SNR is spatially resolved in X-rays and optical.
In addition, we obtained a $\sim5\arcsec$ 
resolution radio image of the SNR. 
These observations provide the highest spatial
resolution imaging of an X-ray/optical/radio SNR
in that galaxy to date. 
The multi-wavelength morphology, X-ray spectrum and 
variability,
and narrow-band optical imagings are consistent
with a SNR.
The SNR is a shell-shaped object with a diameter
of about $10''$ (24 pc). The morphology of the SNR is consistent across
the wavelengths while the \chandra\ spectrum can be well fitted with a
nonequilibrium ionization model with an electron temperature of 2.8 keV and a 
0.3--7 keV luminosity of
$1.6\times10^{37}$ \lum. The age of the SNR is estimated to be $1700-5800$ 
years.

\end{abstract}


\section{Introduction}

Until recently, only supernova remnants (SNRs) in our Galaxy and the
Magellanic Clouds were resolved at X-ray wavelength. However, studies of
Galactic SNRs can be limited by the lack of reliable
distance estimates, high interstellar absorption at optical and X-ray
wavelengths, and the high level of confusion in many Galactic  
fields (Magnier et al. 1995). These limitations are overcome by
studies of SNRs in nearby extragalactic
systems such as the Large and Small Magellanic Clouds (see, e.g., Hughes
2001). More recently, Kong et al. (2002, 2003) and Williams et al. (2004)
discovered five X-ray resolved SNRs in the nearest (780 kpc) spiral
galaxy, M31.
More importantly, these SNRs were also seen at optical and radio wavelengths. 
Such
observations provide a rare opportunity to perform multi-wavelength
morphological studies of SNRs beyond the Magellanic Clouds. 

The question naturally arises whether there are X-ray resolved SNRs in
some nearby dwarf galaxies at a distance between the Magellanic Clouds and
M31. NGC~6822 is one of the nearest dwarf irregular galaxies in the Local
Group beyond the Magellanic Clouds and has a rich history in
observations (see, e.g., Perrine 1922; Hubble 1925).  
At a distance of about 500 kpc (McAlary et al. 1983; Gallart, Aparicio, \&
Vilchez 1996), NGC~6822 is sufficiently close that one can explore its
SNRs with high resolution imaging. However, the relatively high
absorption of NGC~6822 is a substantial problem ($A_B=1.02$; Schlegel
et al. 1998). Despite its proximity, there is only
one known SNR in NGC~6822. 
The SNR Ho~12 was first identified as an
emission-line object by Hodge (1969, 1977) but it was not clear if Ho~12
was an H II region or a SNR. Ho~12 was then confirmed as a
SNR based on narrow-band imaging and
optical spectroscopy (Smith 1975; D'Odorico, Dopita, \& Benvenuti 1980;
D'Odorico, \& Dopita 1983). In the X-ray band, an X-ray source was found
associated with Ho~12 from the   
{\it Einstein} HRI (Markert \& Donahue 1985), and
\rosat\ PSPC (Eskridge \& White 1997, EW97 hereafter). Although both
\einstein\
and \rosat\ observations indicated that the X-ray source is likely to be a
SNR, they
could not rule out that it may be an H II region or an X-ray binary. The
only published radio detection near Ho~12 was from the Very
Large Array (VLA) observations (Dickel et al. 1985). An extended source
with an integrated flux density of
$(1.9\pm2)$ mJy at 1.4 GHz was associated with Ho~12.

In this paper, we report on an archival \chandra\ observation of the
luminous SNR,
Ho~12, in NGC~6822. In addition, we present narrow band optical imagings 
from the Local Group Survey \footnote{http://www.lowell.edu/$\sim$massey/lgsurvey} (LGS; Massey et al. 2001) and
a new radio image
of Ho~12 from combined VLA archival data sets.
In section 2, we detail the observations and
the
results from observations in different wavelengths. A discussion about the
nature of Ho~12 is presented in section 3.

\section{Observations and Results}
\subsection{X-ray Data}

\begin{figure*}[t]
\epsfig{file=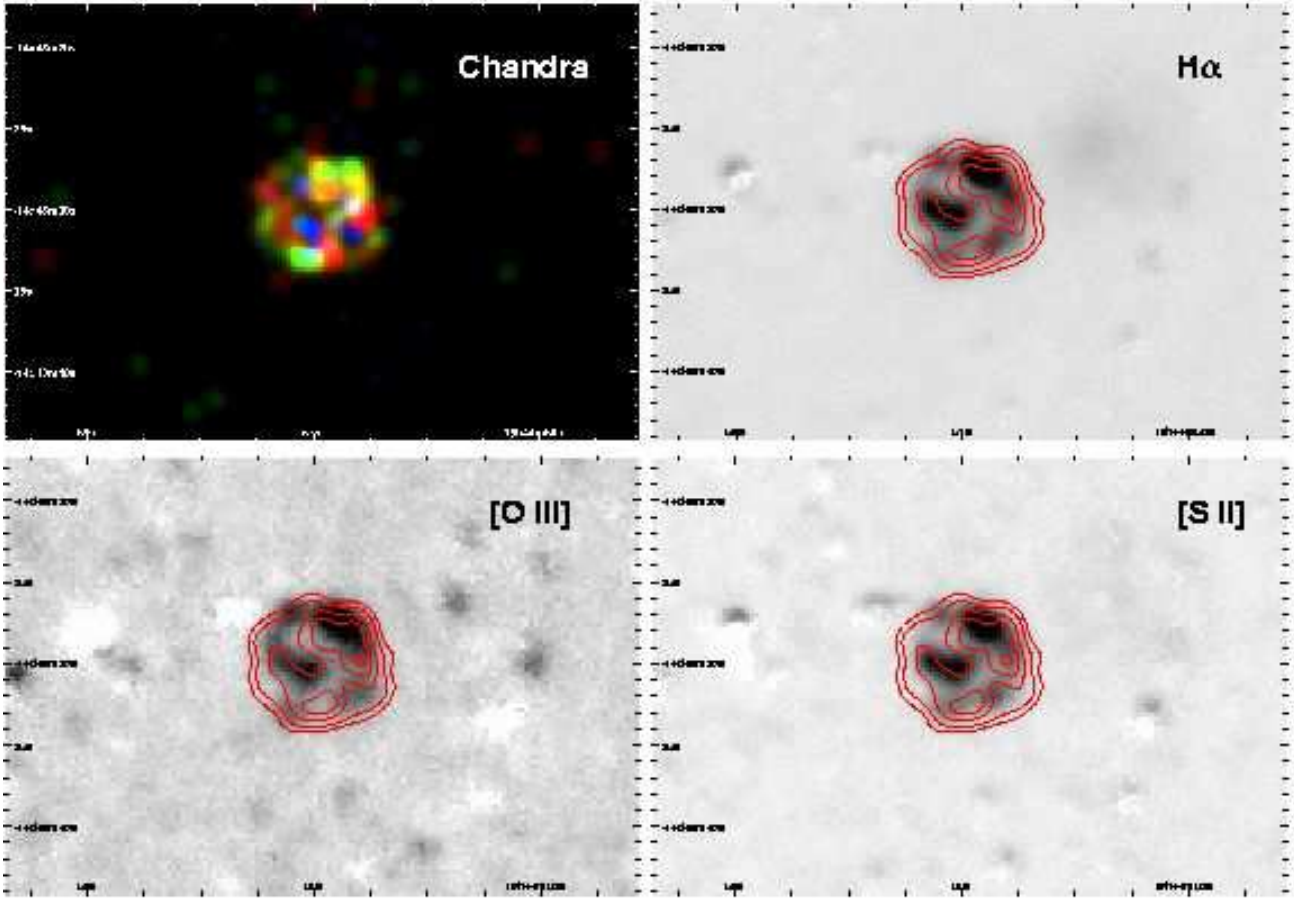,height=12cm}
\caption{True color \chandra\ ACIS-I (upper left) and LGS narrow-band
(H$\alpha$: upper right; [O~III]: bottom
left; [S~II]: bottom right) images of Ho~12.
The \chandra\ image was constructed
from the soft (red: 0.3--9 keV), medium (green: 0.9--1.5 keV), and hard
(blue: 1.5--7 keV) energy bands. The pixel size is $0.248''$, and the
image has been slightly smoothed with a $0.5''\sigma$ Gaussian function.
The diameter of the SNR is about $10''$.
The narrow-band optical images are
continuum-subtracted with the \chandra\
X-ray contours. The contour map was created from a smoothed image with a
$0.5''\sigma$ Gaussian function. Contours are at 10\%, 20\%, 40\%, 60\%,
80\% and 100\% of the peak count of the X-ray SNR. North is up and east is
left.
}
\end{figure*}

NGC~6822 was observed with the \chandra\ Advanced CCD Imaging Spectrometer
(ACIS-I)\footnote{http://asc.harvard.edu/proposer/POG/index.html} on 2002 
November 4 for 28 ksec. The ACIS data were telemetered
in very faint mode (VFAINT).
We reprocessed the X-ray data with the newest calibration (CALDB v2.26)
and at the same time, we screened the raw data in order to utilize the
VFAINT mode, and applied the charge
transfer inefficiency correction to correct the loss of
charge in a CCD as it is shifted from one pixel to the next during
readout\footnote{http://asc.harvard.edu/ciao/threads/acisapplycti/}. 
We also removed the 0.5-pixel
randomization during the pipeline processing to improve the resolution of on-axis {\it Chandra}
source\footnote{http://asc.harvard.edu/ciao/threads/acispixrand/}
(see also Garcia et al. 2001). In order to reduce the instrumental background, we screened the data to allow only photon energies in the range of 0.3--7 keV. We also searched for periods of high
background using source free regions but none was present. We therefore
used all the data for analysis. Data were reduced and analyzed with the
CIAO v3.0.2 package, and spectral analysis was
performed with Sherpa v3.0.

In the \chandra\ image, it is clear that a bright X-ray source is located
at about $10''$ from the aim-point of the observation. The position of 
this X-ray source is consistent with previous detections of Ho~12 
by \einstein\ (Markert \&
Donahue 1985) and \rosat\ (EW97).
To improve the absolute astrometry of the \chandra\ data, we
cross-correlated
detected \chandra\ sources (by WAVDETECT in the CIAO) with the USNO-B1.0 catalog
(Monet et
al. 2003). Within $4'$ of Ho~12, two X-ray sources were less than
$1''$ from bright USNO stars ($B \approx 13$, $B-R \approx 2$); they are
very
likely to be foreground stars. We matched the \chandra\ positions with the
USNO stars and shifted the \chandra\ image by $0.85''$ and $0.53''$ in
right ascension and declination, respectively. After correcting for this
aspect offset, the coordinates of the center of Ho~12 are found to be
R.A.=19$^h$ 44$^m$ 56$^s$.5,
Decl.=$-14^{\circ}$ 48$'$ 30$''$ (J2000.0). The source is clearly extended
with a diameter of $\sim 10''$ ($\sim 24$ pc at 500 kpc). To improve the
spatial
resolution of the image, we applied subpixel event repositioning (SER)
techniques (Li et al. 2004; see also Tsunemi et al. 2001; Li et al. 2003).
SER techniques are based on the premise that
the impact position of events can be refined, according to the
distribution of charge among affected CCD pixels. It is proven that SER 
techniques can improve the spatial resolution by as much as $\sim 60\%$.
We here employed an energy and charge-split dependent SER model on the
\chandra\ image.
All the results reported here are based on the reprocessed image.
Figure 1 shows the smoothed ``true color'' X-ray image of Ho~12. The color
scheme is defined as 0.3--0.9 keV (red), 0.9--1.5 keV (green), and 1.5--7
keV (blue). It is clear that most of the X-ray emission is from below 1.5
keV. 

We extracted the energy spectrum from a   
$5''$ radius and background from an annulus centered on the source; the total 
source counts are 211 while the background
subtracted counts are 205.75. In  
order to allow $\chi^2$ statistics to be used, the spectrum was grouped
into at least 20 counts per spectral bin. The fitting statistic was then used by Sherpa (with CHI DVAR) to find the model parameters that best fit the X-ray spectral data.
Response files were created according to the CCD temperature with
standard CIAO routines; quantum efficiency degradation of ACIS was also
corrected by
using the CALDB v2.26. We fitted the data with several single-component
spectral models including absorbed power law, blackbody,
Raymond-Smith, and nonequilibrium ionization (NEI) models. 
The Raymond-Smith model is a simple collisional equilibrium ionization model, while the NEI model is appropriate for modeling SNRs whose age is smaller than the time required to reach ionization equilibrium. The NEI model consists of an electron temperature ($kT_e$) and an ionization timescale ($n_et$), where $n_e$ and $t$ are the mean electron density and the elapsed time after the plasma was shock heated to a constant temperature $kT_e$. Both models are often applied to study X-ray emission from extragalactic SNRs (see, e.g., Wang 1999; Schlegel, Blair, \& Fesen 2000; Hughes, Hayashi, \& Koyama 1998). 
For the
Raymond-Smith and NEI models, we fixed the chemical
abundances in number of atoms relative to hydrogen (He:
$8\times10^{-2}$, C: $2\times10^{-5}$, N: $4.5\times10^{-6}$, O:
$1\times10^{-4}$, Mg: $2\times10^{-5}$, Si: $4\times10^{-6}$, S:
$1.6\times10^{-6}$) which gave the best
fit from optical spectroscopy (D'Odotico \& Dopita 1983). The results of
the
spectral fits are given in Table 1 and Figure 2. It is clear that except
for the
NEI model, all models did not give a good fit ($\chi^2_{\nu} > 2$). The
best-fit electron temperature of the NEI model is $2.84^{+6.13}_{-2.06}$
keV with
$N_H=(3.0^{+1.9}_{-0.7})\times10^{21}$
cm$^{-2}$. The Galactic absorption for NGC~6822 is 
$9.5\times10^{20}$ cm$^{-2}$, while the observed intrinsic column
associated with this region of NGC~6822 is about $2\times10^{21}$
cm$^{-2}$ (Hodge et al. 1991). Therefore, the absorption derived by our
spectral fit is
consistent with the optical measurement.  
The 0.3--7 keV luminosity is
$1.6\times10^{37}$ ergs s$^{-1}$. 

\begin{table*}
\centering{
\caption{Best-fitting Spectral Parameters}
\begin{tabular}{lcccccc}
\hline
\hline
Model & $N_H$  & $\alpha$ $^a$ & kT & $\log n_e t$ $^b$ & $L_X$ $^c$&
$\chi^2_{\nu}/$dof\\
      & $(\times10^{21}$ cm$^{-2})$ & & (keV) & & &\\
\hline
Power-law & $6.8^{+3.8}_{-2.6}$ & $7.4^{+1.7}_{-2.5}$ &&& 196 &
$2.31/6$\\
Blackbody & $1.0^{+1.9}_{-1.0}$ & & $0.16^{+0.03}_{-0.03}$ & &0.57 &
$2.53/6$ \\
%RS & $1.5^{+1.8}_{-0.9}$ & & $1.13^{+0.33}_{-0.37}$ & &0.93 &
%1.12/6 \\
Raymond-Smith & $7.1^{+4.5}_{-2.5}$ & & $0.38^{+0.21}_{-0.38}$ & & 2.38 &
$5.45/6$ \\
NEI & $3.0^{+1.9}_{-0.7}$ & & $2.84^{+6.13}_{-2.06}$ &
$9.89^{+0.29}_{-0.16}$ & 1.60 & $0.86/5$ \\
\hline  
\end{tabular}
}
\par
\medskip
\hspace{2.7cm}\begin{minipage}{0.8\linewidth}
\footnotesize  
NOTE --- All quoted uncertainties are 90\% confidence.\\
$^a$ Power-law slope.\\
$^b$ Ionization timescale in units of s cm$^{-3}$.\\
$^c$ 0.3--7 keV luminosity ($\times 10^{37}$ ergs s$^{-1}$), assuming a
distance of 500 kpc.
\end{minipage}
\par
\end{table*}

\vspace{0.5cm}
\begin{inlinefigure}
\rotatebox{-90}{\epsfig{file=spectrum.ps,height=8.5cm}}
\caption{\chandra\ ACIS-I spectrum of Ho~12. The data can be fit with an
NEI model with $N_H=3.0\times10^{21}$ cm$^{-2}$ and $kT=2.84$ keV.}
\end{inlinefigure}

\subsection{Optical Data}

We obtained the H$\alpha$, [O~III], [S~II], $V$-band, and $R$-band
images of NGC~6822 from the LGS. These images have been properly flat-fielded and the
geometric distortions removed so that the coordinates in the images
are good to $\sim 0.25''$ and the images at the different bandpasses
are registered with one another. We therefore were easily able to
subtract the $V$-band continuum from the [O~III] image and the
$R$-band continuum from the [S~II] and H$\alpha$ images in order to
make the line-emitting sources stand out.  These images were then
compared to the X-ray images of Ho~12, revealing similar X-ray and
optical sizes and morphologies, as shown in Figures 1 and 3.

To provide convenient optical flux estimates for Ho~12 which
compliment those from the X-ray and radio data, we roughly calibrated
the LGS narrow-band images.  The H$\alpha$ calibration was done using
the bright H~II regions {\em Hubble}~$V$ and {\em Hubble}~$X$ (Hubble 1925).  The
published H$\alpha$ fluxes of these objects are 4.8 and 3.8 $\times
10^{-12}$ ergs cm$^{-2}$ s$^{-1}$, respectively (O'Dell, Hodge \&
Kennicutt 1999). We measured the counts of these H II regions in an
aperture with 30$''$ radius in the star-subtracted LGS H$\alpha$
image.  These measurements provided a rough calibration for the LGS
H$\alpha$ image of 6$\times 10^{-19}$ ergs cm$^{-2}$ s$^{-1}$
ct$^{-1}$.  This rough calibration contains a systematic error because
the $R$-band contains the H$\alpha$; however, the ratio of the
broadband to narrow-band filter widths (1510/80) suggests that the
nebular contamination in the broadband filters should be $\laeq$10\%.

We then measured the counts in the star-subtracted LGS narrow-band
images of Ho~12 using a 4.5$''$ radius aperture.  Applying the
calibration from {\em Hubble}~$V$ and {\em Hubble}~$X$, the total H$\alpha$ flux of Ho~12 is
6.9 $\times 10^{-14}$ ergs cm$^{-2}$ s$^{-1}$. This H$\alpha$ flux is
in good agreement with the surface brightness measured by Killen and
Dufour (1982), which results in a total flux of 8$\times 10^{-14}$
ergs cm$^{-2}$ s$^{-1}$ if a radius of 4$''$ is applied.

The emission-line ratios for Ho~12 have already been spectroscopically
measured by Smith (1975), who found [S~II]/H$\alpha$ = 0.5 and
[O~III]/H$\beta$ = 0.9.  Using his reddening value of $A_V=1.2$, we
infer an observed [O~III]/H$\alpha$ = 0.2.  These ratios provide [S~II]
and [O~III] total flux estimates of 3.5$\times$10$^{-14}$ and
1.5$\times$10$^{-14}$ erg s$^{-1}$ cm$^{-2}$, respectively.  Applying
the measured source counts from Ho 12 in the LGS [S~II] and [O~III]
images, rough calibrations for the LGS [S~II] and [O~III] images of
NGC 6822 are 6.6 $\times 10^{-19}$ ergs cm$^{-2}$ s$^{-1}$ ct$^{-1}$
and 1.2 $\times 10^{-18}$ ergs cm$^{-2}$ s$^{-1}$ ct$^{-1}$,
respectively.  We note that these rough calibrations also contain a
systematic error because the $V$-band contains [O~III] and the
$R$-band contains [S~II]; however, the ratio of the broadband to
narrow-band filter widths for these bandpasses (940/55 and 1510/81 for
$V$/[O~III] and $R$/[S~II], respectively) suggest that the nebular
contamination in the broadband filters should be $\laeq$10\%.

\subsection{Radio Data}

The radio data were obtained from the VLA archive. It includes the
original
Dickel et al. (1985) data set taken in 1984, as well as two longer
observations taken by others for interests other than the SNR in 1985
and 1995. All three data sets were observed in VLA B-array, i.e., with a
spatial
resolution of about 5\arcsec\ at 1.4 GHz. We did not use the 5 GHz
observations included in the data as they had different phase centers and
thus much less sensitivity on the position of the SNR. After some initial
flagging of bad data points, all data sets were reduced with the new
VLARUN
procedure in NRAO's AIPS package. The final image was made using the
combined data after one iteration of self-cal on strong sources in the
field.

The resulting total effective on-source time of the combined 1.4 GHz
observations is about 7 hours, yielding an image noise of about 60
$\mu$Jy/beam. 
However, the archival VLA data sets were taken using different calibrators
and different epochs. In addition, some of the calibrator positions have
been corrected or updated to more acurate positions, resulting in small
uncertainties in the radio positions of each individual archival data set.
Compared to the NRAO VLA Sky Survey
(NVSS\footnote{http://www.cv.nrao.edu/nvss/}, Condon et al.\ 1998), we have
measured an offset of 4.4\arcsec\ in R.A.\ and 0.4\arcsec\ in Dec.\ using seven
bright radio sources in our final image. Given the measured shift, and the
observation that the size and position of the radio source overlap in
coordinates with the optical and X-ray source within one beam, we conclude
that the radio, optical and X-ray source all are from the same remnant, Ho
12. We have shifted the radio data 3.9\arcsec\ Westward to obtain the
overlays shown in Fig.\ 3. With the correction in radio coordinates and
this shift, this suggests that the {\it Chandra}/optical alignment (Sect.\
2.1) and radio reference frame coincide within a fraction of 1\arcsec,
i.e., about one pixel in the {\it Chandra} image.
Also due to combining different data sets, the spatial resolution
($6.3\arcsec\times4.4\arcsec$) and the $7.7\arcsec\times6.3\arcsec$
deconvolved size of the SNR are different from the values measured by
Dickel et al. (1985). The integrated flux of the SNR, 1.57$\pm$0.26 mJy,
is nevertheless consistent. The radio data show a shell-type SNR, with a
brighter region west of the center, and some asymmetric extentions toward
the north and east rims, possibly related to the regions with the brighter
H$\alpha$ knots. Due to the low resolution, there is no clear
correspondence between the radio emission and X-ray knots.

\begin{figure*}
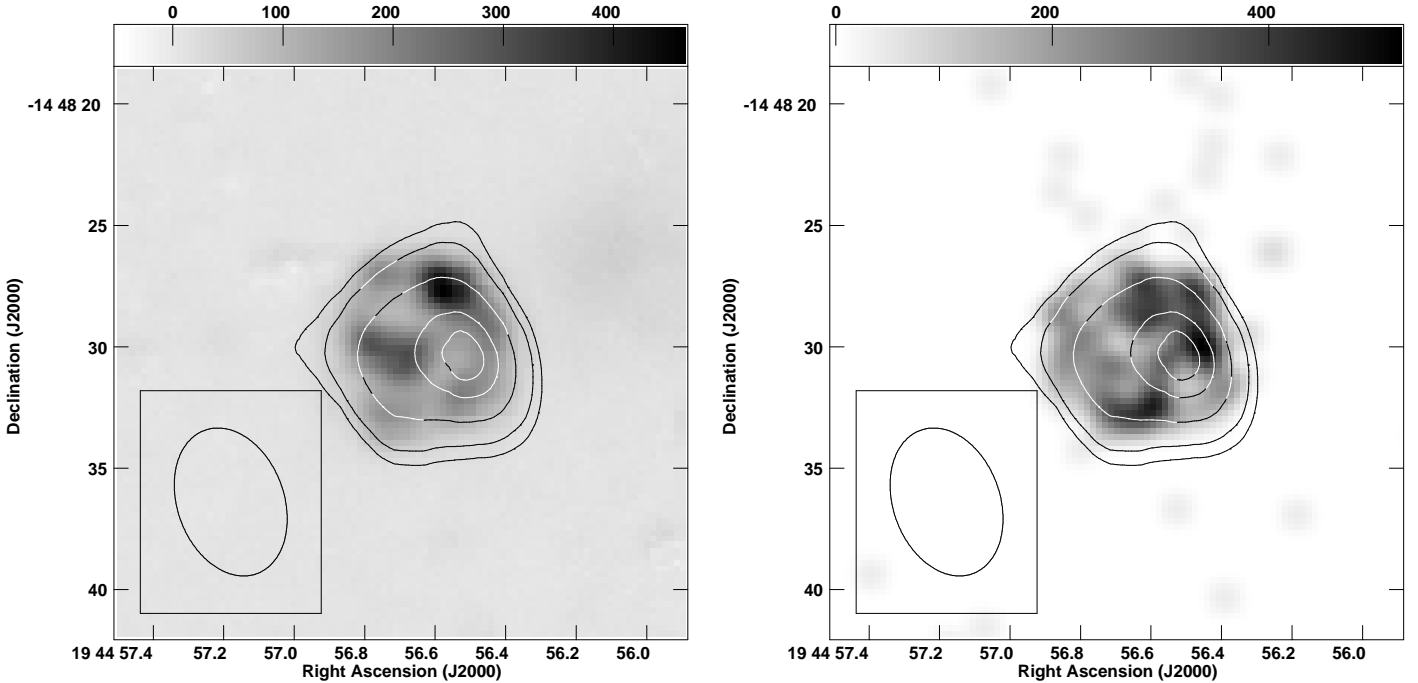

\epsfig{file=f3a.ps,height=9cm} \hfill
\epsfig{file=f3b.ps,height=9cm}
\caption{The 0.25, 0.30, 0.40, 0.50, and 0.55 mJy/beam contours of the
radio detection overlaid on the H$\alpha$ (left) and {\it Chandra}
(right) gray scales. The resolution of the 1.4 GHz radio image is about
5\arcsec, and
the radio image is shifted 3.9\arcsec in R.A.\ to match the optical and
X-ray images.}
\end{figure*}

\section{Discussion}
Ho~12 is clearly resolved with \chandra\ into a shell-shaped object. The
diameter of Ho~12 as measured by \chandra\ is about $10''$, corresponding
to 24 pc. The size and the morphology of the X-ray remnant roughly agrees
with its optical counterpart. It is therefore no doubt that the X-ray
source is the SNR, Ho~12. 
The X-ray spectrum of Ho~12 can be fit with an absorbed NEI model although
the uncertainties of the electron temperature are large due to the low count
rate. Using this model,
the 0.3--7 keV luminosity of the SNR is about $1.6\times10^{37}$ ergs
s$^{-1}$, which is similar to some SNRs in the Magellanic Clouds (e.g.,
Hughes et al. 1998; Williams et al. 1999). The relatively high 
electron temperature
(2.8 keV) is
not unusual; for example, N132D in the Large Magellanic Cloud requires a
two-component NEI model for which the temperatures are 0.8 and 2.7 keV
(Favata et al. 1997). The usual explanation for the hot component is that
it comes from the shock-heated swept-up circumstellar medium or it is due
to the inhomogeneity of the interstellar medium (ISM). Assuming Ho~12 is
in the adiabatic
expansion phase, we can estimate its
physical parameters through the Sedov solution. Using Equations (1) and 
(2) from Kong et al. (2002) and assuming an initial explosion energy of
$3\times10^{50}$ ergs (Blair, Kirshner, \& Chevalier 1981), radius of 12
pc, and shock
temperature of $T_s=2.8^{+6.1}_{-2.0}$, we obtain an age estimate of
$3020^{+2740}_{-1320}$ years and a density estimate of
$n_0=0.03^{+0.06}_{-0.02}$ cm$^{-3}$.

The SNR nature of Ho~12 was called into question due to the X-ray
variability as observed by \einstein\ and \rosat\ (EW97).
However, the \einstein\ HRI count rate quoted by
EW97 was from Fabbiano et al. (1992) in which it
was derived from the entire field of view. This is misleading
since there is another bright source in the field as discussed by
Markert \& Donahue (1985). The count rate of Ho~12 measured by Markert \&
Donahue (1985) is about 33\% of the rate used by EW97. If we take this
into account,
there is indeed no significant variation between the \einstein\ and
\rosat\ observations (see Table 1 of EW97). To compare
the fluxes derived from different instruments, we converted the count
rates into fluxes (0.2--2 keV) by using the best-fit NEI model with 
PIMMS\footnote{http://heasarc.gsfc.nasa.gov/docs/software/tools/pimms.html\#user}.
For the \einstein\ HRI data (taken in 1979), we derived a background
subtracted count rate
of 0.0014 c/s which is consistent with Markert \& Donahue (1985). We also
checked the HRICFA 
database\footnote{http://heasarc.gsfc.nasa.gov/W3Browse/einstein/hricfa.html} 
for which the count rate is corrected for vignetting, deadtime, mirror
scattering, and quantum efficiency, the count rate of Ho~12 is 0.002 c/s
and we used this value for calculating the flux. For the \rosat\ PSPC
(taken in 1992), our measured count rate is 0.011 c/s, consistent with the
PSPC WGA catalog\footnote{http://wgacat.gsfc.nasa.gov}. There are two
\rosat\ HRI datasets (taken in 1995 and 1996; see Zang \& Meurs 2001)
of Ho~12. From the Brera
Multi-scale Wavelet HRI catalog (BMW-HRI; Panzera et al. 2003), Ho~12 was
clearly
detected with count rates of 0.0035 c/s (1995) and 0.0037 c/s (1996).
Our own
measurements of the archival data yield 0.0031 c/s and 0.0035 c/s; it was
underestimated as we did not correct the count rate for the vignetting
and point spread function. We therefore used the values from the BMW-HRI
catalog for analysis.
Using PIMMS, we
derived the history of fluxes (in units of \flux) as follows:
$1.02\times10^{-13}$
(\einstein\ HRI), $1.11\times10^{-13}$ (\rosat\ PSPC),
$9.71\times10^{-14}$, and $1.02\times10^{-13}$ (\rosat\ HRI).
The 0.2--2 keV \chandra\ flux is $9.50\times10^{-14}$\flux. The difference
among them
is less than 20\%. Therefore, we conclude that Ho~12 has not shown
variability in the past two decades.    

At optical wavelengths, Ho~12 is clearly detected as an extended
object in the narrow-band LGS images (see Figure 1).  The
[S~II]/H$\alpha$ ratio ($\sim$0.4; Smith 1975; D'Odorico et al. 1980)
of Ho~12 is typical of SNRs (e.g.  Levenson et al. 1995).  The
reddening-corrected spectroscopic [O~III]/H$\alpha$ ratio is low (0.3;
Smith 1975).  This ratio is about one fourth the ratio predicted for a
typical Type IIa SNR shock in a medium with half the solar O abundance
(Vancura et al. 1992).  The low [O~III]/H$\alpha$ ratio is therefore
reasonable considering the known O abundance of the NGC 6822 ISM
($\sim$25\% solar; Lequeux et al. 1979; Pagel et al. 1980).

Our analysis of archival VLA data provides a tighter
constraint on the integrated 1.4 GHz flux density. Dickel et al. (1985) first 
detected this
source as
an extended ($6.7''\times5.6''$) faint ($1.9\pm2$ mJy at 1.4 GHz) object.
However, the noise of their radio image was dominated by confusion due to 
nearby
very bright objects. Therefore, the error of the flux is comparable with 
the observed value and it cannot be claimed to be a solid detection. We
here obtained a more sensitive radio image by adding more observations.
The resulting flux is $1.57\pm0.26$ mJy at 1.4 GHz, consistent with Dickel
et al. (1985).

The morphology of Ho~12 is very similar across all wavelengths. There is a
hint of brightening in the northwestern region from the narrow-band
optical images and the \chandra\ image. Another ``bright spot''
located at the east side of the remnant in the
optical images, however, corresponds to a breakout region in X-rays.
There is no sinificant difference in morphology among the
three narrow-bands images. In X-ray, although most of the emission is from
below 1.5 keV (see Fig. 1 and 2), it is interesting that the distributions
of the soft (0.3--0.9 keV) and medium (0.9--1.5 keV) photons are slightly
different. The soft band elongates along the northeast-southwest
direction, while the medium band tends to distribute along the
northwest-southeast direction. Although the radio image of Ho~12 is in
low resolution ($5''$), the morphology is consistent with the optical and
X-ray images. In particular, the assymetry of the radio emission toward
the north and east sides may assoicate with the H$\alpha$ knots. 
However, because of the large uncertainty in astrometry and low spatial 
resolution, these spatial coincidences are difficult to determine and 
interpret.
A high resolution radio image and spectral index measurement 
of Ho~12 are therefore required to further investigate the radio emission
and multiwavelength morphology of the remnant.

\begin{acknowledgements}
The authors thank K.\ Johnson and A.\ Seth for sharing their 5~GHz NGC\,6822 
source positions prior to publication.
A.K.H.K. was supported by NASA under GO3-4049X and an LTSA grant
NAG5-10705. B.F.W. acknowledges support from NASA through grant GO-3103X from 
the
Chandra X-Ray Center. The National Radio Astronomy Observatory is a facility 
of the National Science Foundation operated under cooperative agreement by 
Associated Universities, Inc.
\end{acknowledgements}

\end{document}